\documentclass[12pt]{article}
\usepackage{graphicx}
\usepackage{epsf}
\usepackage{cite}

\def\be{\begin{equation}}
\def\ee{\end{equation}}
\def\bea{\begin{eqnarray}}
\def\eea{\end{eqnarray}}

\begin{document}

\thispagestyle{empty}
\vspace*{.5cm}
\noindent
HD-THEP-08-6 \hspace*{\fill} 26 January, 2008\\
\noindent
OUTP-08-02P\hspace*{\fill}revised 25 June, 2009

\vspace*{1.0cm}

\begin{center}
{\Large\bf Inducing the $\mu$ and the $B\mu$ Term\\[.4cm] by the Radion 
and the 5d Chern-Simons Term}
\\[1.5cm]
{\large A.~Hebecker$\,^a$, J.~March-Russell$\,^b$ and R.~Ziegler$\,^c$}
\\[.5cm]
{\it ${}^a$ Institut f\"ur Theoretische Physik, Universit\"at Heidelberg,
Philosophenweg 16 und 19\\ D-69120 Heidelberg, Germany}
\\[.3cm]
{\it ${}^b$ Rudolf Peierls Centre for Theoretical Physics, University
of Oxford, 1 Keble Road\\ Oxford OX1 3NP, UK}
\\[.3cm]
{\it ${}^c$ International School for Advanced Studies (SISSA/ISAS), 
Via Beirut 2-4,\\ 34014 Trieste, Italy}
\\[.4cm]
{\small\tt (\,a.hebecker@thphys.uni-heidelberg.de\,,\,\,}
{\small\tt j.march-russell1@physics.ox.ac.uk}
\\
{\small and}
{\small\tt ziegler@sissa.it\,)}
\\[1.0cm]

{\bf Abstract}\end{center}
\noindent
In 5-dimensional models with gauge-Higgs unification, the $F$-term vacuum
expectation value of the radion provides, in close analogy to the 
Giudice-Masiero 
mechanism, a natural source for the $\mu$ and $B\mu$ term. Both the leading 
order gauge theory lagrangian and the supersymmetric Chern-Simons term 
contain couplings to the radion superfield which can be used for this purpose. 
We analyse the basic features of this mechanism for $\mu$ term generation 
and provide an explicit example, based on a variation of the SU(6) 
gauge-Higgs unification model of Burdman and Nomura. This construction 
contains all the relevant features used in our generic analysis. More 
generally, we expect our mechanism to be relevant to many of the recently 
discussed orbifold GUT models derived from heterotic string theory. This 
provides an interesting way of testing high-scale physics via Higgs mass
patterns accessible at the LHC.

\newpage
\section{Introduction}
The generation of a $\mu$ and $B\mu$ term in the Higgs sector of the 
supersymmetric standard model is one of the critical issues in low-energy 
supersymmetry. While the $\mu$ term alone is responsible for Higgsino 
masses, both terms play a central role in realizing an appropriate scalar 
potential in the Higgs sector, ensuring the spontaneous breaking of the 
electroweak gauge symmetry. Since the $\mu$ term respects supersymmetry, 
one might also formulate the $\mu/B\mu$ term problem by asking why this 
term, which would naturally be either very large or exactly zero, happens 
to be of the same order of magnitude as the soft supersymmetry-breaking 
$B\mu$ term~\cite{Kim:1983dt}. 

The two most popular solutions to this problem are provided by the 
Giudice-Masiero mechanism~\cite{Giudice:1988yz} and the next-to-minimal 
supersymmetric standard model~\cite{Fayet:1974pd}. In the latter, 
the scale of the $\mu$ term is set by the vacuum 
expectation value of the scalar component of an extra uncharged chiral 
superfield. By contrast, in the former the $\mu$ term arises from a term in 
the K\"ahler potential, which mimics a $\mu$ term in the superpotential 
after the non-zero $F$ term of the spurion superfield has absorbed part of the 
superspace integrations. Many variants of these mechanisms as well as
other approaches to the problem have since been considered 
(see~\cite{Barbieri:2007tu} for some recent examples). 

In the present paper, we investigate 5-dimensional models with 
gauge-Higgs unification~\cite{Manton:1979kb}, where the $\mu/B\mu$ term 
problem is solved naturally in a way which is very similar to the 
Giudice-Masiero mechanism. Both these terms as well as the gaugino mass 
term and some of the soft scalar masses are generated at the high scale 
in the interplay of the $F$ term of the radion superfield and the chiral 
compensator of ${\cal N}=1$ supergravity with the quadratic gauge theory 
lagrangian~\cite{choi} (see also~\cite{str}). We point out that the resulting 
high-scale relations are changed significantly by the 5d Chern-Simons term 
which, in particular, induces a non-trivial Higgs scalar potential even in the 
absence of an $F$ term of the chiral compensator. 

At the more fundamental level, our motivation for this work is twofold:
On the one hand, orbifold-GUTs~\cite{orb} are arguably {\it the} modern 
framework for 
grand unification. Within this framework, gauge-Higgs unification receives 
a strong motivation from the requirement of a large top Yukawa coupling. 
Furthermore, it is natural that both the radion superfield~\cite{ky} and 
(after radion stabilization) also the chiral compensator develop an $F$-term 
vacuum expectation value. Thus, all ingredients for our mechanism are 
naturally present and the required terms in the supersymmetric Higgs sector 
arise without any further model building assumptions. 

On the other hand, heterotic orbifold model building has recently produced 
some of the most successful string-theoretic realizations of the 
supersymmetric standard model~\cite{Buchmuller:2005jr} (for earlier related 
work see~\cite{Kobayashi:2004ud}). From this perspective, 
the existence of an intermediate energy scale (one or two orders of 
magnitude below the string scale), at which the world appears to be 
5-dimensional, is also well-motivated~\cite{Hebecker:2004ce}. 
It provides one of the few potential solutions to the 
string-scale/GUT-scale problem. Furthermore, gauge-Higgs unification is again 
a natural ingredient in all constructions where the Higgs fields come from 
the untwisted sector, which is indeed the case in many concrete examples. 

The presence of a $\mu$ term in 5d models with gauge-Higgs unification 
has been noticed early on~\cite{bn}.\footnote{
An
alternative proposal in closely related string-theoretic models appears in 
the last paper of Ref.~\cite{Buchmuller:2005jr}.
} 
The simultaneous generation of 
a $B\mu$ term by the $F$-term vev of the chiral compensator, leading to an 
interesting relations between $\mu$ term, $B\mu$ term and non-holomorphic 
soft Higgs masses, has been pointed out in~\cite{choi}. This relation is
maintained in the presence of a 5d Chern-Simons term, which however changes 
the relation with the gaugino masses. As we already mentioned, the 
Chern-Simons term is crucial in 
situations where the $F$ term of the chiral compensator is small. Although 
such a term is generically present in 5d supersymmetric gauge 
theories~\cite{ims} (see also~\cite{Kuzenko:2006ek}), it affects low-energy 
phenomenology only if some of the scalars of the 5d gauge multiplet develop 
large vacuum expectation values~\cite{Hebecker:2004xx}. This is, however, very 
well motivated in stringy realizations of our scenario, where more than 5 
dimensions are originally present. In most cases, some of these extra 
compact dimensions support non-zero Wilson lines which can, from a 
5d perspective, play the role of the required scalar vacuum expectation 
value. In such situations, the supersymmetric Chern-Simons term is 
parametrically as important for low-energy phenomenology as the quadratic 
lagrangian. 

We finally note that a detailed phenomenological analysis of the proposal 
advocated in the present paper has subsequently appeared 
in~\cite{Brummer:2009ug}. In addition to demonstrating the phenomenological 
viability of our setting, this work was essential for bringing an earlier, 
partially incorrect version of this paper in its present form. We will comment 
on the earlier proposal, its problems and their possible resolutions in
more detail below.\footnote{
We
are indebted to Felix Br\"ummer pointing out the problems of the original 
setting.
}

Our paper is organized as follows: We begin in Sect.~\ref{tm} with the 
discussion of an abelian toy model which shows, in a very direct and 
transparent way, how the quadratic gauge theory lagrangian and the 
Chern-Simons term induce, in their interplay with the radion superfield, 
terms that are structurally similar to the $\mu$ and $B\mu$ term and 
soft supersymmetry breaking masses for the `Higgs field'. 

In Sect.~\ref{nag}, we extend our analysis to the non-abelian case, 
providing in particular a superfield expression for the non-abelian 
supersymmetric Chern-Simons term. The derivation of this term, which we
consider to be a very interesting by-product of our investigation, is 
described in more detail in the Appendix. Applying our formulae to 
a U(6)$\,=\,$SU(6)$\times$U(1) model, where the possibility of gauge-Higgs 
unification is particularly apparent from the decomposition 
${\bf 35}={\bf 24}+{\bf 5}+{\bf\bar{5}}+{\bf 1}$ of the adjoint~\cite{bn}, 
we identify the terms involving the two Higgs superfields, the radion and 
the chiral compensator.

We use our previous results to calculate, in Sect.~\ref{cmb}, $\mu$ and $B\mu$ 
term, as well as soft Higgs scalar masses and gaugino masses. As an 
interesting observation we note that, in the absence of the Chern-Simons
term and of an $F$ term of the chiral compensator, $\mu$ term and soft
scalar masses conspire to ensure an exactly flat scalar potential in the 
Higgs sector. However, once the radion is stabilized, a chiral compensator 
$F$ term generically develops and this flatness is lifted. 

In Sect.~\ref{csp}, we give the complete expressions for the $\mu$ term 
and the soft parameters of the gauge-Higgs sector, including the effects 
of the Chern-Simons term and chiral compensator. We then briefly discuss 
the viability of this high-scale input for low-energy phenomenology after 
the renormalization group running down to the electroweak scale. We also 
comment on the influence of the squark masses and trilinear terms on this 
running and on the partially model-dependent high-scale origin of these 
terms (especially in the top quark sector) in our 5d gauge-Higgs unification 
scenario. 

Finally, we provide in Sect.~\ref{bns} an explicit phenomenologically viable 
construction that has all the qualitative features which we used in our 
previous discussion. Our model is closely related to a 5d model for gauge-Higgs
unification by Burdman and Nomura~\cite{bn}. We obtain our model by 
lifting this previous construction to 6 dimensions, where the compact space 
has the topology of a pillow case, and taking a different 5d limit of this 
geometry. In this way the non-zero 5d vev of the scalar component of the gauge 
multiplet is automatically enforced. The rather intricate realization of 
matter fields and Yukawa couplings can essentially be copied from the 
construction of Burdman and Nomura. 

Our summary and conclusions are given in Sect.~\ref{conc}.

\section{The basic mechanism in an abelian toy model}\label{tm}
The supersymmetric 5d U(1) gauge theory has a well-known description in 
terms of a 4d real superfield $V$ and a chiral superfield $\Phi=\Sigma+ 
iA_5+\cdots$, both depending on the extra parameter $x^5$. Using this 
language, the quadratic 5d lagrangian reads~\cite{Marcus:1983wb,
Arkani-Hamed:2001tb}
\be
{\cal L}_2=\frac{1}{4g_5^2}\left[\int d^2\theta\,\,W^2+{\rm h.c.}+
\int d^4\theta\,\left(2\partial_5V-(\Phi+\bar{\Phi})\right)^2\right]\,,
\label{l2}
\ee
where $W$ is the supersymmetric field strength defined in terms of $V$. 
The supersymmetric Chern-Simons term which will in general be present
in this theory takes the form~\cite{Arkani-Hamed:2001tb}\footnote{
Note
that we find a different sign for the second term than is reported 
in~\cite{Arkani-Hamed:2001tb}.
}
\bea
{\cal L}_{cs} & = & c\,\left[\,\int d^2\theta\,\,\Phi\,W^2 + {\rm h.c.} 
\right. 
\nonumber \\ 
& & \hspace*{.5cm} + \left. \frac{2}{3} \int d^4\theta \; \left( \partial_5 
V D_\alpha V - V D_\alpha \partial_5 V  \right) W^\alpha + {\rm h.c.} \right. 
\nonumber \\
& & \hspace*{.5cm} - \left. \frac{1}{6} \int d^4\theta \; \left( 2 \partial_5 
V - (\Phi + \bar{\Phi}) \right)^3 \right]\,.\label{lcs}
\eea

We are interested in the 4d effective field theory obtained after $S^1$ 
compactification of the above model, in particular in the couplings to the 
radion superfield. The following discussion can be viewed as a mild 
generalization of~\cite{Marti:2001iw} (because of the Chern-Simons term) or 
as a significantly simplified version of the derivation of related formulae 
in~\cite{PaccettiCorreia:2004ri}. 

The relevant 4d lagrangian is found by 
simply dropping all terms involving $x^5$ derivatives, replacing $V$ and 
$\Phi$ by their ($x^5$-independent) zero modes, and integrating the result 
over $x^5$. In the rigid case, the latter amounts to a multiplication by 
$L=2\pi R$. By contrast, in the case where the original model is coupled to 
5d supergravity, this multiplicative factor has to be replaced by the radion 
superfield $T$ (or $\bar{T}$) in the holomorphic (antiholomorphic) terms of 
Eqs.~(\ref{l2}) and (\ref{lcs}) and by $(T+\bar{T})/2$ in the $d^4\theta$ 
terms. Here the 4d chiral superfield $T$ is normalized such that
\be
T=L+iB_5\,\,,
\ee
where $B_M$ ($M=0\ldots 3,5$) is the graviphoton of the 5d supergravity 
multiplet. Its pure-derivative coupling in the component action enforces 
the use of the combination $T+\bar{T}$ in the $d^4\theta$ terms in 
Eqs.~(\ref{l2}) and (\ref{lcs}).

However, this is not the only way in which $T$ enters the 4d effective 
theory. From the fact that $\Phi$ contains the gauge field component $A_5$,
and $A_5$ covariantizes the derivative operator $\partial/\partial x^5$, it
follows that the whole superfield has to scale as the inverse size of the 
compact dimension. Thus, we have to perform the replacements 
\be
\Phi\,\to\,\frac{L_0}{T}\Phi\qquad\mbox{and}\qquad \Phi\,\to\,
\frac{2L_0}{T+\bar{T}}\Phi\label{resc}
\ee
in the $d^2\theta$ and $d^4\theta$ terms above. Here we have introduced 
an arbitrary constant $L_0$ with the dimension of length to insure that
the new superfield $\Phi$ has the dimension of mass.

To summarize, the 4d low-energy lagrangian follows from Eqs.~(\ref{l2}) and 
(\ref{lcs}) after suppressing any $x^5$ dependence, multiplying the 
appropriate terms by $T$, $\bar{T}$ or $(T+\bar{T})/2$, and performing the
redefinition of Eq.~(\ref{resc}). The results are
\be
{\cal L}_{2,\,4d}=\frac{1}{4g_5^2}\left[\int d^2\theta\,\,T\,W^2+{\rm h.c.}+
2L_0^2\int d^4\theta\,\frac{(\Phi+\bar{\Phi})^2}{T+\bar{T}}\right]\label{l24}
\ee
and
\be
{\cal L}_{cs,\,4d} = c\,\left[\,L_0\,\int d^2\theta\,\,\Phi\,W^2 + {\rm h.c.} 
+ \frac{4L_0^3}{6} \int d^4\theta\,\frac{(\Phi + \bar{\Phi})^3}
{(T+\bar{T})^2} \right]\,.\label{lcs4}
\ee
To check that the $T$ dependence obtained in this intuitive approach is 
indeed correct, one can work out the component form of the above superfield 
expressions and match it (with appropriate field redefinitions and 
keeping track of all factors $g_{55}$) to the 5d component 
action~\cite{ziegler}. 

Our main point concerning the generation of certain MSSM operators can 
now easily be made. Recall that we want to think of $V$ as containing the 
Standard model gauge multiplet and of $\Phi$ as the Higgs 
superfield.\footnote{
Of 
course, in this simple U(1) toy model $\Phi$ is not charged and the second
Higgs multiplet is missing, but that is irrelevant for now.
} 
If the radion auxiliary field $F_T$ develops a non-zero expectation value, 
it is immediately clear that the superspace integrals in Eq.~(\ref{l24}) 
induce operators
\be
\sim F_TW^2\Big|_1\quad,\quad 
\sim|F_T|^2\Phi^2\Big|_1 \quad\,\quad 
\sim\bar{F}_T\Phi^2\Big|_{\theta^2}\quad,\quad
\sim|F_T|^2\Phi\bar{\Phi}\Big|_1\quad\mbox{and}\quad 
\sim\bar{F}_T\Phi\bar{\Phi}\Big|_{\theta^2}\,.\label{ops}
\ee
The first of them provides gaugino masses, which is often referred to as 
radion mediation~\cite{ky}. The second, which clearly has the structure of 
the MSSM $\mu$ term, provides Higgsino masses.\footnote{
This
can be understood from a slightly different perspective as follows: 
Non-vanishing $F_T$ is the 4d manifestation of an $SU(2)_R$ symmetry twist
of in the 5d background. The latter induces gaugino masses and, since the
Higgsinos are 5d gauginos in the present setting, non-vanishing Higgsino 
masses are also induced~\cite{bn}.
}
Furthermore, both the second and the remaining operators in Eq.~(\ref{ops}) 
contribute to the scalar potential, thereby apparently inducing 
a $B\mu$ term and soft scalar masses in the Higgs sector. However, a more 
careful analysis of Eq.~(\ref{l24}) reveals that all these contributions 
exactly cancel and the scalar potential remains flat. (This fact, which can 
also be understood from a structural perspective~\cite{Barbieri:1985wq}, 
remains true in the non-abelian case.) 

To lift the flatness of the potential and to induce a non-zero $B\mu$ 
term and soft scalar masses in the present 
framework, the effect of the chiral compensator of ${\cal N}=1$ supergravity,
$\varphi=1+F_\varphi\theta^2$, has to be taken into account. More 
specifically, a factor $\varphi\bar{\varphi}$ has to be included the last 
term in Eq.~(\ref{l24}). If $F_\varphi$ develops a non-zero vacuum 
expectation value, operators analogous to those displayed in Eq.~(\ref{ops}) 
(but with one or both of the factors $F_T$ and $\bar{F}_T$ replaced by 
$F_\varphi$ and $\bar{F}_\varphi$) are induced. The total scalar potential 
looses its flatness, which can be described by a non-vanishing $B\mu$ and 
soft scalar mass terms.

If the lowest component of $\Phi$ develops a vacuum expectation 
value, then the Chern-Simons lagrangian of Eq.~(\ref{lcs4}) corrects the 
quadratic order lagrangian of Eq.~(\ref{l24}). Moreover, if $c={\cal O}(1)$ 
and $\langle \Phi\rangle\sim 1/g_5^2$ (both of which are natural values,
as will become clear in the following), these contributions are not 
parametrically suppressed relative to those of Eq.~(\ref{l24}). Thus, gaugino 
masses, $\mu$ and $B\mu$ term, and the Higgs sector soft scalar masses 
are induced on the basis of the fundamental lagrangian of Eqs.~(\ref{l2}) and 
(\ref{lcs}) after coupling it to supergravity and allowing for vacuum 
expectation values of $\Phi$, $F_T$ and $F_\varphi$. As we will explain in 
more detail below, in higher-dimensional unified models an interesting and 
realistic phenomenology can emerge on the basis of this very generic 
mechanism.

\section{Non-abelian generalization}\label{nag}
The ${\cal N}=1$ superfield action of the 5d non-abelian gauge theory~\cite{ 
Marcus:1983wb,Arkani-Hamed:2001tb} can be given in a manifestly 
super-gauge-invariant form using the super-gauge-covariant $x^5$ 
derivative~\cite{Hebecker:2001ke}
\be
\nabla_5=\partial_5+\Phi\,.
\ee
It reads
\be
{\cal L}_2=\frac{1}{2g_5^2}\mbox{tr}\left[\int d^2\theta\,\,W^2+{\rm h.c.}+
\int d^4\theta\,\left(e^{-2V}\nabla_5e^{2V}\right)^2\right]\,,
\label{l2na}
\ee
where the action of $\Phi$ on $e^{2V}$ follows from the standard gauge 
transformation properties of $e^{2V}$, i.e.,
\be
\nabla_5e^{2V}=\partial_5e^{2V}-\Phi^\dagger e^{2V}-e^{2V}\Phi\,.
\ee
For the non-abelian supersymmetric Chern-Simons term we have, unfortunately, 
not been able to derive an equally elegant superfield formula. However,
sacrificing manifest super gauge invariance by restricting ourselves to
Wess-Zumino gauge, the following expression can be derived~\cite{ziegler} 
(see Appendix):
\bea
{\cal L}_{cs} & = & c\,\mbox{tr}\left[\,\int d^2\theta\,\,\Phi\,W^2 + 
{\rm h.c.} \right. 
\nonumber \\ 
& & \hspace*{1cm} + \left. \frac{1}{3} \int d^4\theta \; \left( \{\partial_5 
V,D_\alpha V\} - \{V,D_\alpha\partial_5 V\}\right) W^\alpha + {\rm h.c.} 
\right. 
\nonumber \\ 
& & \hspace*{1cm} - \left. \frac{1}{12} \int d^4\theta \; \left( \{\partial_5 
V,D_\alpha V\} - \{V,D_\alpha\partial_5 V\}\right) W_{(2)}^\alpha) + 
{\rm h.c.} 
\right. 
\nonumber \\
& & \hspace*{1cm} - \left. \frac{1}{6} \int d^4\theta \; 
\left(e^{-2V}\nabla_5e^{2V}\right)^3 \right]\,.\label{lcsna}
\eea
Here curly brackets are used for anticommutators and $W_{(2)}^\alpha$ 
represents the part of $W^\alpha$ which is quadratic in $V$ (recall
that, in Wess-Zumino gauge, $W$ is the sum of a linear and quadratic piece 
in $V$).

Starting from Eqs.~(\ref{l2na}) and (\ref{lcsna}), which are the 
non-abelian generalizations of Eqs.~(\ref{l2}) and (\ref{lcs}), the coupling 
of the radion superfield to the zero modes of the compactified theory can 
be derived in complete analogy to Sect.~\ref{tm}. To recapitulate, one 
simply has to suppress any $x^5$ dependence, multiply the appropriate terms 
by $T$, $\bar{T}$ or $(T+\bar{T})/2$, and perform a redefinition analogous
to that of Eq.~(\ref{resc}). The results are
\be
{\cal L}_{2,\,4d}=\frac{1}{2g_5^2}\mbox{tr}\left[\int d^2\theta\,\,T\,W^2+
{\rm h.c.}+2L_0^2\int d^4\theta\,\frac{(\Phi+\bar{\Phi})^2}{T+\bar{T}}\right]
\label{l2na4}
\ee
and
\be
{\cal L}_{cs,\,4d} = c\,\mbox{tr}\left[\,L_0\,\int d^2\theta\,\,\Phi\,W^2 + 
{\rm h.c.} + \frac{4L_0^3}{6} \int d^4\theta\,\frac{(\Phi + \bar{\Phi})^3}
{(T+\bar{T})^2} \right]\,.\label{lcsna4}
\ee
Clearly, this could have also been obtained by starting from Eqs.~(\ref{l24})
and (\ref{lcs4}), promoting the superfields $V$ and $\Phi$ to appropriate
matrices and introducing the corresponding trace operations. In this 
sense, our above discussion of the 5d superfield expression for the 
non-abelian Chern-Simons term is included merely for completeness (and 
possible other applications). The phenomenology-oriented analysis 
following from now on is based entirely on Eqs.~(\ref{l2na4}) and 
(\ref{lcsna4}), which are straightforward generalizations of Eqs.~(\ref{l24})
and~(\ref{lcs4}). 

We can now be more specific about how we envisage the $\mu$ and $B\mu$ term
generation to proceed in models of this type. To be concrete, let $V$ and 
$\Phi$ take values in the Lie algebra of the GUT gauge group 
U(6)$\,=\,$SU(6)$\times$U(1). Furthermore, let the theory be compactified 
to 4d on an interval such that SU(6) is broken to SU(5)$\times$U(1)$'$ and 
the U(1) is completely broken. In the corresponding decomposition 
of the adjoint representation,
\be
{\bf 35}={\bf 24}+{\bf 5}+{\bf \bar{5}}+{\bf 1}\,,
\ee
we find, as parts of the superfield $\Phi$, the Higgs multiplets $H_u$ and 
$H_d$ in the ${\bf 5}$ and ${\bf \bar{5}}$ of SU(5). The further breaking 
of SU(5) to the standard model gauge group, which could for example also 
be realized by boundary conditions, is not important at the moment. 

Thus, the second term of Eq.~(\ref{l2na4}) gives rise to the following 
contribution to the 4d Higgs lagrangian:
\be
{\cal L}_{2,\,4d}\supset \frac{1}{g_4^2}\int d^4\theta\frac{2L_0\varphi
\bar{\varphi}}{T+\bar{T}}(H_u+\bar{H}_d)(H_d+\bar{H}_u)\,.\label{lh2}
\ee
Furthermore, if $\Phi$ develops a vev $\langle\Phi\rangle=v\,\mbox{\bf 1}$, 
consistent with the assumed boundary-breaking of the U(1)\footnote{
In 
an earlier version of this paper, a $\Phi$-vev $\sim\mbox{diag}(1,1,1,1,1,-5)$
inside the adjoint of SU(6) was assumed. This is inconsistent with an orbifold 
breaking of SU(6) to SU(5)$\times$U(1)$'$. The desired breaking by boundary 
conditions can nevertheless be realized, e.g. by introducing a brane localized 
adjoint superfield and giving it a large vev $\sim\mbox{diag}(1,1,1,1,1,-5)$. 
However, the bulk vev of $\Phi$ induces a bulk mass for the ${\bf 5}$ and 
${\bf \bar{5}}$ Higgs fields. This is easy to see since the gauge symmetry 
is broken in the 5d bulk. Hence the `broken' $A_5$ components, which form 
some of the Higgs scalars, become massive in 5d. Equivalently, when thinking 
at the zero-mode level of a corresponding $S^1$ compactification, this mass
term must be present since the ${\bf 5}$ and ${\bf \bar{5}}$ chiral 
multiplets become part of the massive vector multiplet. On an interval with
boundary-breaking, massless 4d fields in these representations nevertheless 
survive since only a certain 
linear combination of the bulk and brane ${\bf 5}$ and ${\bf \bar{5}}$ fields 
is `eaten' by the vector multiplets which become massive in the breaking of
SU(6) to SU(5)$\times$U(1)$'$. However, these massless Higgs fields now have 
a non-trivial bulk profile because of their bulk mass. This profile depends 
on the size of the $\Phi$-vev and affects both the calculation of soft terms 
and of Yukawa couplings, thereby significantly complicating the subsequent 
analysis. This set of problems as well as its resolution by simply using U(6) 
instead of SU(6) was pointed out to us by Felix Br\"ummer (see 
also~\cite{Brummer:2009ug}). 

We also note that U(6) is, of course, a product gauge group allowing for 
independent coefficients of the SU(6)- and U(1)-kinetic terms as well as of
the CS terms of SU(6), U(1) and of the mixed CS terms (see e.g.~\cite{
Kim:2008kn}). Since, given the above $\Phi$-vev, only the mixed CS term is 
relevant for our analysis, we do not complicate our notation by making all 
those independent coeffcients explicit.
}, 
the second term of Eq.~(\ref{lcsna4}) gives rise to the following correction 
to this lagrangian 
(up to quadratic order):
\be
{\cal L}_{cs\,,4d}\supset 2cL_0v\int d^4\theta\frac{(2L_0)^2\varphi
\bar{\varphi}}{(T+\bar{T})^2}(H_u+\bar{H}_d)(H_d+\bar{H}_u)\,.\label{lhcs}
\ee
Here we have assumed that, with the exception of $H_u$ and $H_d$, all 
the zero-mode components of the chiral adjoint $\Phi$ have been eliminated
by orbifolding (or acquired a large mass in another way). Note also that,
since we are not interested in the dynamics of $T$ and $\varphi$ at the 
moment, we have suppressed the constant term $\sim v^3$ in Eq.~(\ref{lhcs}). 
A term $\sim v^2$, which would have to be linear in $H_u$ and $H_d$, does 
obviously not arise for group theoretic reasons. 

In a vacuum where $T$ and $\varphi$ develop non-zero $F$ terms, 
Eqs.~(\ref{lh2}) and (\ref{lhcs}) provide, in addition to the kinetic terms 
for the Higgs multiplets, $\mu$ term, $B\mu$ term and soft scalar masses in
the Higgs sector. The relevant operators are analogous to those given 
explicitly in the case of our abelian toy model in Eq.~(\ref{ops}) of the 
previous section. In addition, the first terms of both Eq.~(\ref{l2na4}) 
and (\ref{lcsna4}) contribute to the standard model gauge kinetic term and 
to the corresponding gaugino masses. We devote the following two sections 
to the discussion of the resulting SUSY breaking pattern.

\section{Calculating the $\mu$ and $B\mu$ term and the Higgs-sector soft 
scalar masses}\label{cmb}
To begin, we ignore the possible Chern-Simons term and focus on the 
phenomenological implications of Eq.~(\ref{lh2}) and the first term 
of Eq.~(\ref{l2na4}). We assume the existence of a (meta-)stable 
almost-Minkowski vacuum in which Re$\,T=L_0$ and both $F_T$ and $F_\varphi$ 
have non-zero values. Using the chiral compensator approach to supergravity, 
the scalar potential in the Higgs sector (with canonical 4d field 
normalization) is easily obtained: We simply have to integrate out the 
auxiliary-field vectors $F_{H_u}$ and $F_{H_d}$ on the basis of 
Eq.~(\ref{lh2}) while treating $T$, $F_T$ and $F_\varphi$ as fixed external 
sources. The result reads
\be
{\cal L}_{4,\,can.}\supset -\left(|F_\varphi|^2-\frac{F_\varphi\bar{F}_T+
\bar{F}_\varphi F_T}{T+\bar{T}}\right)(H_u+\bar{H}_d)(H_d+\bar{H}_u)\,.
\label{pot}
\ee
We emphasize that, in contrast to the last section, in this and the 
following equations $H_u$ and $H_d$ are the scalar components of the 
corresponding superfields and their normalization has been modified to 
make the 4d kinetic term canonical. The corresponding Higgsino mass term
can be directly read off from Eq.~(\ref{lh2}):
\be
{\cal L}_{4,\,can.}\supset -\left(\bar{F}_\varphi-\frac{\bar{F}_T}{T+\bar{T}}
\right)\lambda_u\lambda_d+\mbox{h.c.}\,\label{higgsino}
\ee
where $\lambda_u$ and $\lambda_d$ are two-component Weyl spinors. This 
determines the value of the $\mu$ parameter, which is conventionally 
defined as the coefficient of the Higgsino bilinear:
\be
\mu=\bar{F}_\varphi-\frac{\bar{F}_T}{T+\bar{T}}\,.
\ee
Similarly to the gaugino mass 
\be
m_{1/2}=\frac{\bar{F}_T}{T+\bar{T}}\,,
\ee
a non-zero $\mu$ parameter arises as a consequence of $F_T$, even if 
$F_\varphi$ vanishes. 

Furthermore, if the Higgs scalar potential is parameterized by (see 
e.g.~\cite{Martin:1997ns})
\be
{\cal L}_{4,\,can.}\supset -(|\mu|^2+m_{H_u}^2)|H_u|^2-(|\mu|^2+
m_{H_d}^2)|H_d|^2-(B\mu)H_uH_d+\mbox{h.c.}+\mbox{quart.~terms}\,,
\ee
we read off from Eq.~(\ref{pot}) that $B\mu$, $m_{H_u}$ and $m_{H_d}$ are 
given by (see also~\cite{choi})
\be
B\mu=|\mu|^2+m_{H_u}^2=|\mu|^2+m_{H_d}^2=|F_\varphi|^2-
\frac{F_\varphi\bar{F}_T+\bar{F}_\varphi F_T}{T+\bar{T}}\,.\label{spar}
\ee
In contrast to the $\mu$ parameter, these scalar mass parameters vanish 
if $F_\varphi=0$. This is a result of the very specific generalized no-scale 
structure of the superfield expression in Eq.~(\ref{l2na4}). In terms of the 
conventional parameterization of the component lagrangian with soft terms,
it implies a somewhat surprising exact cancellation between $|\mu|^2$ and
$m_{H_u}^2$ as well as between $|\mu|^2$ and $m_{H_d}^2$ in 
Eq.~(\ref{spar}). Clearly, the phenomenological implications of the above
formulae crucially depend on the values of $F_T$ and $F_\varphi$ (especially 
on their relative size), on which we now briefly comment.

At the tree level, the compactification of 5d supergravity on $S^1/Z_2$ or
$S^1/(Z_2\times Z_2')$ gives rise to a K\"ahler potential of no-scale
type for the radion,
\be
K_0(T,\bar{T})=-3\ln(T+\bar{T})\,.
\ee
An effective constant superpotential can be introduced if the boundary 
conditions at the two ends of the interval preserve different ${\cal N}=1$
subalgebras of the original ${\cal N}=2$ SUSY. (Alternatively, the same 
effect can arise as a result of some non-perturbative boundary effect, such
as brane gaugino condensation.) In the resulting no-scale model, 
supersymmetry is broken by $F_T$, but $T$ remains a flat direction. At the 
same time, $F_\varphi$ remains exactly zero. For our purposes, this 
approximation (in the case that this is a reasonable approximation to the 
physical vacuum) is insufficient since, as already mentioned in 
Sect.~\ref{tm}, the Higgs sector scalar potential remains exactly flat in 
this case. 

Thus, we have to take the stabilization of the radion $T$ seriously from 
the very beginning and to determine $F_T$ and $F_\varphi$ in the context 
of a stabilized vacuum. It is well-known that $F_\varphi$ is generically 
non-zero in such situations (implying, in our context, that a Higgs sector 
scalar potential will be generated). 

Starting from the no-scale situation described above, stabilization of $T$ 
can arise as a result of either K\"ahler corrections or $T$-dependent 
superpotential terms. To be as generic as possible, we assume a model where,
on the basis of a corrected K\"ahler potential and superpotential,
\be
K(T,\bar{T})=K_0(T,\bar{T})+\Delta K(T,\bar{T})\qquad\mbox{and}\qquad W(T)\,,
\ee
a (meta-)stable almost-Minkowski vacuum is produced (see e.g.~\cite{ 
Luty:1999cz,Luty:2002hj}). The equations of motion for $F_T$ and $F_\varphi$ 
(and thus their vacuum values) can be obtained on the basis of the flat-space 
superfield lagrangian
\be
\int d^4\theta \varphi\bar{\varphi}\Omega(T,\bar{T})+\int d^2\theta\varphi^3
W(T)+\mbox{h.c.}\,,\label{tflag}
\ee
where $\Omega=-3\exp(-K/3)$ is the so-called `superspace kinetic 
energy'~\cite{wb}. 

For the purpose of this paper, we do not want to specify a stabilization 
mechanism for $T$ and extremize Eq.~(\ref{tflag}) explicitly. Instead, we
restrict ourselves to deriving a simple relation between the $F$ terms 
of the radion and the chiral compensator. This can be achieved rather
easily: First, assume that Eq.~(\ref{tflag}) possesses a SUSY-breaking 
minimum with vanishing cosmological constant. In this minimum, $W$ takes 
some vacuum expectation value $W_0$. We now go to a different K\"ahler-Weyl 
frame, defined by the requirement that the superpotential $W'$ in this 
frame is constant, $W'=W_0$. Such a change of frames can be viewed as a 
redefinition of the chiral compensator. The new chiral compensator $\varphi'$ 
is defined in terms of $T$ and $\varphi$ by
\be
W(T)\varphi^3=W'\varphi'^3\,.
\ee
In this new frame, $F_{\varphi'}=0$, which is an immediate consequence of 
vanishing vacuum energy and constant superpotential (see 
e.g.~\cite{Hebecker:2004sb}). Thus,
\be
\varphi=\varphi'\cdot\left(\frac{W(T)}{W_0}\right)^{-1/3}=1\cdot\left(1+
\frac{W_TF_T\theta^2}{W_0}\right)^{-1/3}\,.
\ee
To lighten notation, we can now suppress the index `0' of $W$ and simply 
conclude that
\be
F_\varphi=-\frac{W_T}{3W}F_T
\ee
in the physical vacuum. This formula allows for a simple evaluation of 
the previously derived supersymmetric and SUSY-breaking Higgs mass terms 
and their relation to gaugino masses in any concrete model of radius 
stabilization. Note that, for a generic function $W(T)$, we expect 
$F_\varphi\sim F_T/T$ on dimensional grounds. This relation is also found 
in the specific model of~\cite{Luty:1999cz}. The SUSY-breaking effects of 
$F_\varphi$ and $F_T$ are then parametrically equally important.

\section{Including the effect of the Chern-Simons term and some 
phenomenological consequences}\label{csp}
We now repeat the analysis of the previous section on the basis of the 
complete lagrangian of Eqs.~(\ref{lh2}) and (\ref{lhcs}). Integrating out 
$F_{H_u}$ and $F_{H_d}$, the following (canonically normalized) scalar 
potential arises:
\be
{\cal L}_{4,\,can.}\supset-\left[|F_\varphi|^2-\frac{(F_\varphi\bar{F}_T+
\mbox{h.c.})}{T+\bar{T}}\,\frac{1+2c'}{1+c'}+\frac{|F_T|^2}{(T+\bar{T})^2}
\,\frac{2c'^2}{(1+c')^2}\right](H_u+\bar{H}_d)(H_d+\bar{H}_u)
\,,\label{potcs}
\ee
where
\be
c'=2cvg_5^2\,.\label{cp}
\ee
Note that the no-scale argument ensuring the complete flatness of the 
scalar potential in the absence of $F_\varphi$ has broken down. The reason 
is as follows: While the Chern-Simons term by itself respects the generalized 
no-scale structure, the presence of a fixed vev $v$ breaks this structure. 
For this it is crucial that the vev is truly fixed in the sense that no 
corresponding fluctuations are allowed - a situation which indeed arises 
in certain orbifold models (see below). 

Similarly, the Higgsino mass term, Eq.~(\ref{higgsino}), is now replaced by 
an analogous expression following from Eqs.~(\ref{lh2}) and (\ref{lhcs}):
\be
{\cal L}_{4,\,can.}\supset \left(\bar{F}_\varphi-\frac{\bar{F}_T}{T+\bar{T}}
\,\frac{1+2c'}{1+c'}\right)\lambda_u\lambda_d+\mbox{h.c.}
\ee
The gaugino mass is also affected by the Chern-Simons term. Although 
$F_\Phi$ does not develop a vacuum expectation value, the first term in 
Eq.~(\ref{lcsna4}) affects the normalization of the gauge kinetic term 
and hence the gaugino mass. Thus, we can summarize all effects by giving 
the following set of SUSY-breaking parameters and the $\mu$ term:
\bea
m_{1/2}&=&\frac{\bar{F}_T}{T+\bar{T}}\,\frac{1}{1+c'}\,,\label{m12}\\ 
\nonumber\\
B\mu&=&|\mu|^2+m_{H_u}^2=|\mu|^2+m_{H_d}^2\\
&=&|F_\varphi|^2-\frac{(F_\varphi\bar{F}_T+\mbox{h.c.})}{T+\bar{T}}\,
\frac{1+2c'}{1+c'}+\frac{|F_T|^2}{(T+\bar{T})^2}\,\frac{2c'^2}{(1+c')^2}\,,
\nonumber\\ \nonumber\\
\mu&=&\bar{F}_\varphi-\frac{\bar{F}_T}{T+\bar{T}}\,\frac{1+2c'}{1+c'}\,.
\label{mut}
\eea
The most striking feature of this result is, as without the Chern-Simons term, 
the equality between the $B\mu$ term and the parameters $|\mu|^2+m_{H_u}^2$ 
and $|\mu|^2+m_{H_d}^2$~\cite{choi}. We now briefly discuss the 
phenomenological consequences of this relation:

It is a well-known fact (see e.g.~\cite{Martin:1997ns}) that electroweak 
symmetry breaking, i.e. the destabilization of the vacuum with vanishing 
Higgs expectation values, requires
\be
(B\mu)^2>(|\mu|^2+m_{H_u}^2)(|\mu|^2+m_{H_d}^2)\,.\label{lar}
\ee
At the same time, positivity of the quadratic part of the scalar potential 
along the $D$-flat directions is guaranteed if 
\be
2(B\mu)<(|\mu|^2+m_{H_u}^2)+(|\mu|^2+m_{H_d}^2)\,.\label{sma}
\ee
For the parameters that we have found, both inequalities turn into 
equalities, apparently disfavouring our scenario phenomenologically. 
However, our previous analysis was performed at a high scale (the GUT scale 
or the orbifold-GUT compactification scale, which is usually only 
marginally lower). Thus, our findings are, in fact, very encouraging since 
even small running effects can easily turn the high-scale equalities into the 
desired inequalities of Eqs.~(\ref{lar}) and (\ref{sma}). 

We now discuss in more detail how this running modification of our 
high-scale relations may occur. The crucial renormalization group 
equations are
\bea
16\pi^2\frac{d}{dt}\mu&=&\mu\left[3|y_t|^2-3g_2^2\right]\,,\\ \nonumber\\
16\pi^2\frac{d}{dt}(B\mu)&=&B\mu\left[3|y_t|^2-3g_2^2\right]
+\mu\left[6a_t\bar{y}_t+6g_2^2M_2\right]\,,\\ \nonumber\\
16\pi^2\frac{d}{dt}m_{H_u}^2&=&6|y_t|^2\left[m_{H_u}^2+m_{Q_3}^2+m_{u_3}^2
\right]+6|a_t|^2-6g_2^2|M_2|^2\,,\\ \nonumber\\
16\pi^2\frac{d}{dt}m_{H_d}^2&=&-6g_2^2|M_2|^2\,,
\eea
where, except for writing $B\mu$ instead of $b$, we follow the conventions 
of~\cite{Martin:1997ns}. Since, for the purposes of this paper, we are only 
interested in qualitative features, we have neglected all Yukawa couplings 
and trilinear couplings (except those of the top) as well as the U(1) gauge 
coupling $g_1$. 

From the above equations we first immediately recognize the well-known fact 
that, starting with $m_{H_u}^2=m_{H_d}^2$ at a high scale, one generically 
finds $m_{H_u}^2<m_{H_d}^2$ at the electroweak scale, essentially because of 
the effects of the large top Yukawa coupling. We also see from the formulae 
at the beginning of this section that both $m_{H_u}^2$ and $m_{H_d}^2$ can 
easily be negative from the beginning in our setting.

Thus, $(|\mu|^2+m_{H_u}^2)<(|\mu|^2+m_{H_d}^2)$ at the low scale and the 
inequalities of Eqs.~(\ref{lar}) and (\ref{sma}) can, in principle, be 
satisfied simultaneously. Clearly, whether this actually happens depends on 
the running of $\mu$ and $B\mu$ and on their initial values. This depends, 
in turn, on the fundamental parameters $F_T$, $F_\varphi$ and $c'$ of our 
construction. Furthermore, the running also depends on the soft masses
and trilinear couplings in the top quark sector. Since, as we will discuss in 
more detail in Sect.~\ref{bns}, the matter fields originate in bulk 
hypermultiplets, the relevant terms come from the superfields 
expressions~\cite{Marti:2001iw}
\be 
{\cal L}_{hyp.,\,4d}\supset \int d^4\theta \varphi\bar{\varphi}\frac{1}{2}
(T+\bar{T})\left(H^\dagger e^{-2V}H+H^ce^{2V}H^{c\dagger}\right)+
\int d^2\theta\varphi^3H^c\Phi H+\mbox{h.c.}
\ee
Unfortunately, as will again be explained in Sect.~\ref{bns} referring to 
the model of~\cite{bn}, realistic Yukawa couplings require many such 
hypermultiplet terms with non-trivial bulk profiles as well mixing with 
brane localized charged fields. Thus, we can not simply write down the 
soft squark masses and trilinear couplings without entering more deeply 
in the matter sector of our model. 

Nevertheless, we see from the above that, using the freedom of choosing 
$F_T$, $F_\varphi$, $c'$ and of the bulk field localization and bulk-brane 
mixing in the matter sector, it is very plausible that realistic low-scale 
SUSY-breaking parameters and $\mu$ term can result from our fundamental 
high-scale formulae, Eqs.~(\ref{m12})--(\ref{mut}). In situations without 
a Chern-Simons term, a numerical analysis of the running of the relevant 
parameters has already been performed in Ref.~\cite{choi}, using certain 
plausible assumptions about soft parameters in the top-quark sector. The 
authors came to the conclusion that, given the strong high-scale constraints, 
correct electroweak symmetry breaking is difficult to achieve. They identified 
the prediction $m_{H_u,\,H_d}^2=-m_{1/2}^2$ as one of the main reasons for 
this difficulty. However, in our model with a Chern-Simons term, precisely 
this constraint is lifted. In fact, as one can see from 
Eqs.~(\ref{m12})--(\ref{mut}), the parameters $m_{1/2}^2$ and $m_{H_u,\,H_d}^2$
blow up for different negative values of $c'$, implying that any 
high-scale ratio of these quantities can, in principle, be realized. Indeed,
as has recently been demonstrated in~\cite{Brummer:2009ug}, the inclusion of 
the Chern-Simons term in this type of gauge-Higgs unification models allows 
for a realistic low-energy phenomenology.

\section{An explicit SU(6) orbifold-GUT model}\label{bns}
Both the U(6) model analysed above as well as the more minimal pure SU(6)
model briefly discussed in a footnote in Sect.~{\ref{nag}}, do not represent
`clean' versions of field-theoretic orbifolding. Indeed, the U(1) factor in
U(6) does not allow, in the presence of charged matter, for a breaking by 
a $Z_2$ symmetry of the original action. The pure SU(6) model, on the other
hand, inherently relies on the gauge symmetry breaking by (non-orbifold)
boundary conditions. Thus, it is interesting to see whether a 5d model can be 
found which realizes all the essential features of our scenario by just 
modding out a set of $Z_2$ symmetries. In the present section, we provide a 
positive answer to this question, modifying the model of~\cite{bn}
appropriately. However, this construction has problems of its own which 
are related to precision gauge coupling unification (see below). 

Although we are ultimately interested in 5d orbifold GUT models with 
gauge-Higgs unification, the simplest way to approach our model is from a 
6d perspective. We start from 6d ${\cal N}=2$ super-Yang-Mills theory 
with gauge group SU(6) compactified on a torus $T^2$. The torus is 
parameterized by a complex coordinate $z$ with the fundamental domain being 
defined by $0\le\mbox{Re}z<2\pi R_6$ and $0\le\mbox{Im}z<2\pi R_5$. We 
restrict the field space of the model by requiring invariance under two 
orbifold projections $P$ and $P'$. With each of these operations we associate 
SU(6) matrices which characterize the orbifold action in gauge space and which 
we denote by the same symbol: $P=i\,\mbox{diag}(1,1,1,1,1,-1)$ and $P'=
\mbox{diag}(1,1,-1,-1,-1,-1)$. The invariance requirements for the 
${\cal N}=1$ vector superfield $V$ contained in the 6d gauge multiplet are
\be
PV(z)P^{-1}=V(-z)\qquad\mbox{and}\qquad P'V(z-\pi/2)P'^{-1}=V(-(z-\pi/2))\,.
\ee
Similar relations, but with an extra minus sign, hold for the chiral 
superfield $\Phi$, which contains the remaining degrees of freedom of the 
6d gauge multiplet. 

\begin{figure}
\begin{center}
\includegraphics[width=10cm]{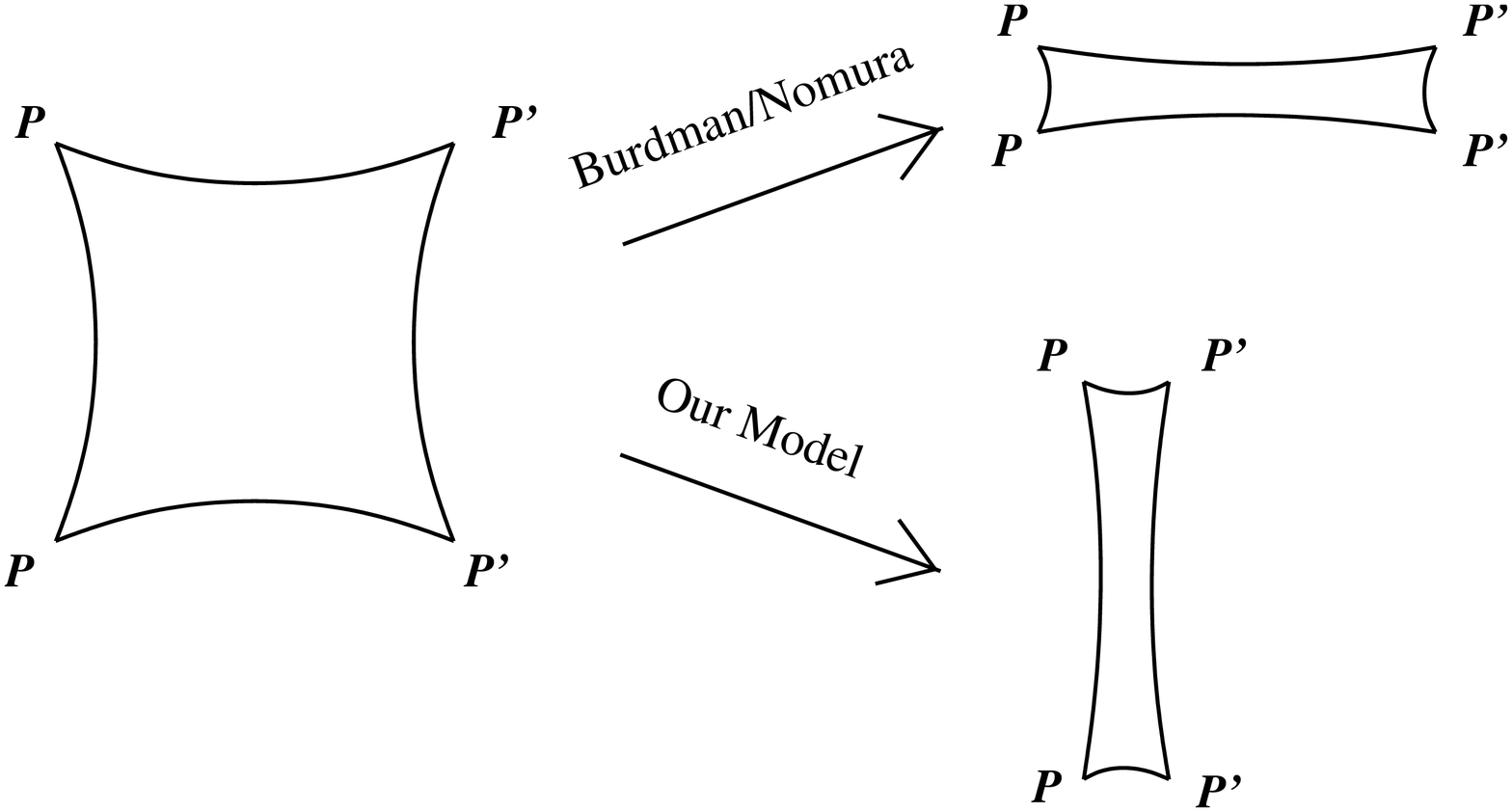}
\caption{In two different 5d limits, the 6d model described in the text 
goes over into the model of Burdman/Nomura or into `Our Model'.}
\label{6d}
\end{center}
\end{figure}

The resulting theory can be visualized as a 6d model the compactification 
space of which has the geometry of a pillow (cf.~Fig.~\ref{6d}). This space 
has four conical singularities, each with deficit angle $\pi$, two of which 
are due to the projection $P$ and the other two of which are due to the 
projection $P'$. Correspondingly, the gauge symmetry is locally restricted 
at these singularities to SU(5)$\times$U(1) for $P$ and to SU(4)$\times$SU(2)$
\times$U(1) for $P'$. 

We now observe that by taking the limit $R_5\to 0$, we arrive precisely at 
the 5d orbifold GUT model with gauge-Higgs unification of Burdman and 
Nomura~\cite{bn}. This limit is illustrated in Fig.~\ref{6d}. Indeed, 
in this limit the pillow degenerates to an interval and the fixed points 
with gauge group SU(5)$\times$U(1) (labelled by $P$) merge into a boundary 
of the 5d space with the same local gauge symmetry. Analogously, the two 
fixed points with gauge group SU(4)$\times$SU(2)$\times$U(1) merge and 
play the role of the other boundary or brane. 

We define our model by keeping $R_5$ finite and taking the limit $R_6\to 0$. 
This situation, which is also visualized in the figure, corresponds 
again to a 5d model compactified on an interval. However, the two
boundaries are now equivalent and the gauge symmetry at the boundary, which 
is restricted by both $P$ and $P'$, is the intersection of the two groups 
left invariant by the two projections. It is just the gauge symmetry of the 
standard model plus an extra U(1) factor (the U(1) left over when SU(6) is 
broken to SU(5)). 

The model that we have thus obtained is similar but not identical to the 5d 
model of Sect.~\ref{nag}: The original gauge symmetry, which is SU(6) rather 
than U(6), is broken at each boundary of the interval to ${\cal G_{\rm SM}}
\times$U(1) rather than simply to SU(5)$\times$U(1). In addition, the vacuum 
expectation value of $\Phi$ takes a less symmetric form. To determine this 
vacuum expectation value, 
we first recall that the scalar part of the chiral superfield $\Phi$ (which 
we denote by the same symbol) reads $\Phi=A_6+iA_5$ in the 6d construction. 
Furthermore, if a charged particle encircles the stretched pillow (labelled
`Our model' in Fig.~\ref{6d}) in the short direction, it experiences a gauge 
rotation
\be 
P\cdot P'=\exp[i(\pi/4)T]=\exp\left[i\int_0^{\pi R_6}A_6dx^6\right]\,. 
\ee
Here $T=\mbox{diag}(1,1,-1,-1,-1,1)$ is the generator of the gauge twist 
$P\cdot P'$ which is felt in the bulk of our effective 5d space and which 
breaks SU(6) to SU(3)$\times$SU(3)$\times$U(1). Thus, after dimensional 
reduction from 6d to 5d, we find $\langle\Phi\rangle=v\,\mbox{diag}(1,1,-1,
-1,-1,1)$ with $v=1/(4R_6)$. 

This result may appear puzzling since it seems to imply that the physical 
effects of $v$, introduced via the Chern-Simons term, become dominant in the 
5d limit $R_6\to 0$. However, this is not the case for the following reason: 
The smallest $R_6$ for which our 6d motivation of the 5d model makes sense is 
$R_6\sim g_6$. For smaller $R_6$, the 6d approach is compromised by the fact 
that the strong-coupling scale of the 6d gauge theory lies below the 
compactification scale. Through the relation 
$1/g_5^2\sim R_6/g_6^2$, this limiting situation gives rise to an effective 
5d gauge-coupling $g_5\sim\sqrt{R_6}$. We thus conclude from Eq.~(\ref{cp})
that the dimensionless parameter $c'$ governing the size of the physical 
effects induced by $v$ is indeed ${\cal O}(1)$ if the coefficient of the 
Chern-Simons term in the original lagrangian is $c\sim{\cal O}(1)$. Of course, 
the 6d supersymmetric gauge theory does not allow for a Chern-Simons term.
However, the 5d theory obtained after $S^1$-compactification includes such a
term because of loop effects. The group-theoretic structure of these loop
induced Chern-Simons terms, which have been discussed in some detail in
Sect.~5 of~\cite{Hebecker:2004xx} (see also~\cite{ims}), is somewhat 
different from that of the tree-level 5d Chern-Simons term.\footnote{
Such 
structures are possible because the loop-induced prepotential does not 
have to be holomorphic at the origin, $\Phi=0$. This allows for gauge 
invariant expressions different from tr$\,\Phi^3$.
}
However, the 
coefficient follows entirely from group theory and matter content and 
is thus naturally ${\cal O}(1)$. We will not derive these terms explicitly 
in the present 6d-motivated model but only reiterate that, as we claimed 
before, the physical effects of the Chern-Simons term in the presence of 
$v$ do indeed arise in more fundamental constructions and are, in general, 
comparable to the effects derived from the quadratic lagrangian. 

Let us finally turn to the problem of standard model matter fields and 
Yukawa couplings in the presented gauge-Higgs unification model. This is, 
in principle, a highly non-trivial issue since charged hypermultiplets 
have to be introduced in the bulk in such a way that, after the orbifold 
projections, the correct low-energy spectrum results. Furthermore, large 
4d Yukawa couplings (in particular that of the top quark) can only result
from bulk gauge couplings because the two Higgs doublets come from the 
chiral superfield $\Phi$ in the $\bf 35$, which is part of the gauge 
multiplet and can not have any other interactions in the 5d (or 6d) bulk. 

However, concerning all of these issues we can simply refer the reader to 
the 5d SU(6) model of~\cite{bn}. In this model, all of the above issues 
have been solved: For example, the down- and up-type quarks are introduced 
as hypermultiplets in the ${\bf 15}$ and ${\bf 20}$ of SU(6) in the bulk, 
which mix with extra 4d chiral superfields introduced on the branes. It has 
then been shown that the top- and other Yukawa couplings can be correctly 
reproduced from the 5d couplings with the gauge multiplet. A similar procedure 
works for the leptons. The hierarchies of the Yukawa couplings can be 
realized by allowing for 5d bulk masses for the hypermultiplets, which lead
to exponential profiles of the fields and hence to very different effective 
4d couplings for the zero modes of the hypermultiplets. 

Indeed, the whole construction of~\cite{bn} can straightforwardly be lifted 
to 6 dimensions. The field content in 5d and 6d is exactly the same. The 
orbifold $S^1/(Z_2\times Z_2')$ can be replaced by $T^2/(Z_2\times Z_2')$, 
as is visualized in Fig.~\ref{6d}. Instead of placing extra 4d chiral 
superfields and 4d superpotentials on the boundaries of the 5d interval, 
those can equally well be placed at the conical singularities of the 6d
orbifold. In short, the whole construction goes through without change. 
A critical issue appears to be the introduction of 5d bulk masses for the 
hypermultiplets, which is not possible for charged hypermultiplets in 6
dimensions. However, the 6d hypermultiplets may be charged under extra U(1) 
gauge groups. Wilson lines of these gauge groups (i.e. vacuum expectation 
values of $A_6$) then play the same role as 5d bulk masses and lead to 
localization effects for the zero modes. To summarize, we could simply copy 
the relevant pages of~\cite{bn}, changing the language from 5d to 6d. We 
will not do so since, in this paper, we do not intend to go beyond the 
demonstration that the type of model underlying our discussion of SUSY 
breaking in the Higgs sector does indeed arise in phenomenologically viable 
GUT models. 

Although the 6d lift of the 5d model 
of~\cite{bn} and its `opposite' 5d limit appear to be a very nice motivation
of our 5d framework, this is not the only way to approach our construction. 
Instead, we could simply say that our model is defined, from the start, 
on a 5d interval with gauge group SU(6) in the bulk. At each boundary, the 
gauge group is broken to ${\cal G}_{\rm SM}\times$U(1) (which is not a
$Z_2$ orbifold breaking) and a non-zero vacuum expectation value for 
$\Sigma$ is enforced by the boundary conditions. The inclusion of matter and 
the generation of Yukawa couplings can be achieved in analogy to the 
similar 5d gauge-Higgs unification model of~\cite{bn}. From this perspective, 
our model remains 5-dimensional. The `pillow' of Fig.~\ref{6d} and its 5d 
limit merely serve to convince the reader that non-orbifold 5d boundary 
conditions are natural, for example as the result of two merging conical 
singularities with gauge breaking by $P$ and $P'$. 

We finally note that, since the 5d vev used in this section does not preserve 
the SU(5) subgroup, large threshold corrections to gauge-coupling unification 
will generically be present~\cite{Hebecker:2004xx}. This is not necessarily 
fatal since the size of these thresholds and the way in which they affect the 
low-energy couplings is highly model dependent. However, it would 
require a more detailed analysis to establish whether a fully realistic 
low-energy phenomenology can emerge. Such an analysis is beyond the scope
of the present investigation.

\section{Conclusions}\label{conc}
We have analysed supersymmetry breaking and the supersymmetric $\mu$ term 
in the Higgs sector of 5-dimensional models with gauge-Higgs unification. 
This setting is well-motivated both from the perspective of 5d or 6d 
orbifold GUTs, which are arguably the simplest realistic grand unified 
theories on the market, as well as from the perspective of the most 
successful heterotic string models. 

Gaugino masses, soft Higgs masses, as well as the $\mu$ and $B\mu$ term 
are generated in a natural way once the $F$ terms of the radion superfield 
and the chiral compensator acquire non-zero vacuum expectation values. This 
happens in many of the simplest models where the radion (the size of the 
5th dimension) is stabilized with the help of a non-trivial superpotential. 
The relative size of the SUSY-breaking parameters and the $\mu$ term 
depend on ratio of the two $F$ terms, $F_\varphi/F_T$. The overall scale is 
set by the ratio of the radion $F$ term and the size of the extra dimension, 
$F_T/T$. This means that low-scale supersymmetry is realized if the 
high-scale theory exhibits weak Scherk-Schwarz breaking (known as 
radion mediation). 

In addition to the effects based on the quadratic gauge theory lagrangian, 
the 5d supersymmetric Chern-Simons term can play a crucial role. This is, in 
fact, expected since the Chern-Simons term is an unavoidable part of generic 
5d models compactified on an interval. Its importance for the low-energy 
effective theory depends on the presence of a large vacuum expectation value 
of the 5d scalar in the gauge multiplet. Such a vacuum expectation value 
can be viewed as a Wilson line from the perspective an underlying 6d or 
string model. Its size is then naturally of the right order of magnitude to 
compete with the effects of the quadratic lagrangian. 

If, as explained above, supersymmetry breaking is governed by both the 
quadratic lagrangian and the Chern-Simons term, all relevant terms are 
generated just on the basis of the $F$ term of the chiral compensator. 
One can then consider the limit where the $F$ term of the chiral compensator 
vanishes, corresponding e.g. to the stabilization of the radion purely by 
K\"ahler corrections. 

The details of the resulting low-energy phenomenology are sensitive 
to the various high-scale parameters, in particular $F_\varphi$, $F_T$ and 
the vacuum expectation value of the 5d scalar (the real part of the chiral 
adjoint). However, an interesting feature that appears to be universal 
within the class of models that we have investigated is the high-scale 
relation $B\mu=|\mu|^2+m_{H_u}^2=|\mu|^2+m_{H_d}^2$. This relation between 
$B\mu$ term, $\mu$ term and soft Higgs masses is at the borderline of validity 
of the standard inequalities which have to be imposed for successful 
electroweak
symmetry breaking. Thus, we rely on running effects to lift the equality 
$m_{H_u}^2=m_{H_d}^2$, which is standard, and on an appropriate running of
$\mu$ and $B\mu$ to satisfy the necessary low-energy constraints. As 
demonstrated in~\cite{Brummer:2009ug}, the Chern-Simons term, which lifts 
certain extra constraints, is crucial to avoid the negative conclusions 
concerning the low-energy phenomenology of related models reached 
in~\cite{choi}. Thus, the proposed version of supersymmetric gauge-Higgs 
unification with a 5d Chern-Simons term defines an interesting new class of 
potentially realistic GUT models.

\section*{Acknowledgments}
We would like to thank Felix Br\"ummer for pointing out a problem in an
earlier version of this paper, as well as its resolution.

\section*{Appendix}
This appendix is devoted to the construction of a superfield expression 
for the non-abelian supersymmetric Chern-Simons term. Suppressing a possible 
overall prefactor, the superfield expression for the abelian 5d Chern-Simons 
term is given by~\cite{Arkani-Hamed:2001tb}
\bea
{\cal L}_{cs} & = & \int d^2\theta\,\,\Phi\,W^2 + {\rm h.c.} 
\nonumber \\ 
& & \hspace*{-.7cm} + \frac{2}{3} \int d^4\theta \; \left( \partial_5 
V D_\alpha V - V D_\alpha \partial_5 V  \right) W^\alpha + {\rm h.c.}  
\nonumber \\
& & \hspace*{-.7cm} - \frac{1}{6} \int d^4\theta \; \left( 2 \partial_5 
V - (\Phi + \bar{\Phi}) \right)^3 .\label{lcsz}
\eea
The simple 4d procedure for the non-abelian generalization, i.e. the 
replacement $V \to e^{\pm2V}$, does not work in this case. Instead, we 
construct the non-abelian lagrangian by matching an appropriate 
superfield expression (in Wess-Zumino gauge) to the component action. 
Working within this approach is straightforward because the number of 
possible superfield actions is highly restricted and the calculation can be 
performed in close analogy to the abelian case.

Our starting point is the 5d Chern-Simons action of the (non-supersymmetric)
non-abelian gauge theory, which can be constructed from the 5d Chern-Simons 
form given in~\cite{Nakahara:1990th}:
\be
\label{nacscomp}
{\cal L}_{cs\, gauge} = \epsilon^{MNOPQ} \, {\rm tr} \left({\frac{1}{4} 
A_M F_{NO} F_{PQ} - \frac{i}{4} A_M A_N A_O F_{PQ} - \frac{1}{10} 
A_M A_N A_O A_P A_Q}\right)
\ee
with the non-abelian field strength 
\be
F_{MN}=\partial_M A_N - \partial_N A_M + i [A_M,A_N]\,. 
\ee
This expression must be reproduced by a superfield lagrangian which contains 
the fields $\Phi,V,W_{\alpha}$ with bosonic components
\bea
\Phi & = & \Sigma(y) + iA_5(y) + \theta^2 F_\Phi(y) \nonumber \\
V_{{\rm WZ}} & = & - \theta \, \sigma^\mu \bar{\theta} A_\mu(x) + 
\frac{1}{2} \theta^2 \bar{\theta}^2 D(x) \label{cexp}\\
W_\alpha & = & \theta_\alpha D(y) - i \left( \sigma^{\mu \nu} 
\right)_\alpha{}^\beta \theta_\beta F_{\mu \nu}(y)\,, \nonumber 
\eea
where $y = x + i \theta \sigma \bar{\theta}$. Note that the field strength 
superfield 
\be
W_\alpha=- \frac{1}{8} \bar{D}^2 \left( e^{-2V} D_\alpha e^{2V} \right)
\ee
gives, in Wess-Zumino gauge, only terms linear and quadratic in $V$:
\be
W_\alpha=W_\alpha^{(1)} + W_\alpha^{(2)}
\ee
with
\bea
W_\alpha^{(1)} & = &  -\frac{1}{4}\bar{D}^2D_\alpha V \hspace*{.8cm}=\,\, 
\theta_\alpha D(y) - 2i \left( \sigma^{\mu \nu} \right)_\alpha{}^\beta 
\theta_\beta \partial_\mu A_\nu(y) 
\nonumber \\
W_\alpha^{(2)} & = & -\frac{1}{4}\bar{D}^2[D_\alpha V,V]\,\,=\,\,2 \left( 
\sigma^{\mu \nu} \right)_\alpha{}^\beta \theta_\beta A_\mu(y) A_\nu(y)\,,
\eea 
which reproduces the expression in Eq.~(\ref{cexp}). 

It is convenient to rewrite Eq.~(\ref{nacscomp}) as
\be
\label {nacscomp5}
{\cal L}_{cs\,gauge} = \epsilon^{\mu \nu \rho \sigma}\,{\rm tr} \left( 
\frac{3}{4} A_5 F_{\mu \nu} F_{\rho \sigma} -\frac{1}{2} \{A_\mu,
\partial_5 A_\nu\} F_{\rho \sigma} + \frac{i}{4} \{A_\mu,\partial_5 A_\nu\} 
A_\rho A_\sigma \right)\,,
\ee
where the curly brackets denote anticommutators. It can be checked that the 
variation of this expression under gauge transformations is a total 
derivative.

The first term in Eq.~(\ref{nacscomp5}) is obtained from a superfield 
lagrangian which is of the same form as in the abelian case:
\be
{\rm tr} \left( \int {d^2 \theta} \, \Phi \, W^\alpha W_\alpha + {\rm h.c.} 
\right)\,.\label{p1}
\ee
The second term is reproduced by a piece which is also similar to the abelian 
case:
\be
{\rm tr} \left( \int {d^4 \theta} \, \left( \{\partial_5 V,D_\alpha V\} - 
\{V,\partial_5 D_\alpha V\} \right) W^\alpha + {\rm h.c.} \right). \label{p2}
\ee
For the last term, it is necessary to use just the part of $W_\alpha$ 
quadratic in $V$:
\be
{\rm tr} \left( \int {d^4 \theta} \; \left( \{\partial_5 V,D_\alpha V\} - 
\{V,\partial_5 D_\alpha V\} \right) W^{\alpha}_{(2)} + {\rm h.c.} \right)\,.
\label{p3}
\ee
The above three terms already reproduce the non-supersymmetric 5d CS term of 
Eq.~(\ref {nacscomp5}), but 5d Lorentz invariance is violated by a term 
$\sim \Sigma F_{\mu\nu}F^{\mu\nu}$ coming from Eq.~(\ref{p1}). This can be 
cured by adding a further contribution, which is a simple 
generalization of the last term in the abelian CS action:
\be
{\rm tr} \int {d^4 \theta}  \left( e^{-2V} \nabla_5 \, e^{2V} \right)^3\,.
\label{p4}
\ee
Here we have used the super gauge covariant derivative 
\be
\nabla_5 \equiv \partial_5 + \Phi,
\ee
acting on $e^{2V}$ as 
\be
\nabla_5 e^{2V} = \partial_5 e^{2V} - \Phi^\dagger e^{2V} - e^{2V} \Phi.
\ee

The relative prefactors of the four contributions of 
Eqs.~(\ref{p1})--(\ref{p4}) are fixed by an explicit calculation and found 
to be consistent with those of the abelian action. Up to an overall 
constant factor, the result is that of~Eq.~(\ref{lcsna}). 
Although the evaluation of this manifestly supersymmetric expression in WZ 
gauge reproduces the CS component lagrangian of Eq.~(\ref{nacscomp}), we were 
not able to show that it transforms into a total derivative under super 
gauge transformations. Most probably this is due to missing extra terms that 
vanish in WZ gauge. It would be interesting to construct these missing 
contributions and achieve manifest super gauge invariance (as it is realized 
for the leading order lagrangian in Eq.~(\ref{l2na})). 

It requires a certain amount of work to extract even just the bosonic part 
of our full superfield Chern-Simons lagrangian. One has to integrate by 
parts using the fact that $\Sigma$ vanishes at the boundaries. Furthermore, 
$F_\Phi$ is set to zero by the equations of motion, while $D$ takes the value
\be
D = -\partial_5 \Sigma + i [\Sigma,A_5]\,.
\ee
The final result is
\be
{\cal L}_{cs} \supset c\left[\,\frac{2}{3} {\cal L}_{cs\,gauge}  - {\rm tr} 
\left( \Sigma F_{MN} F^{MN} + 2 \, \Sigma (D_M \Sigma)\,(D^M \Sigma)\right)
\,\right]\,,
\ee
where
\be
D_M \Sigma = \partial_M \Sigma + i [A_M,\Sigma]\,.
\ee
This also fixes the normalization of our superfield expression relative to 
the non-supersymmetric Chern-Simons term.

\end{document}